%% file: MAIN.tex
\documentclass[astrosymb, twocolumn]{aastex631}
\usepackage{subfigure}
\usepackage{rotating}
\usepackage{mathtools}
\usepackage{xspace}
\usepackage{float} 
\usepackage[nodayofweek,level]{datetime} 

\shorttitle{X-ray Analysis of AGN from the  GALEX Time Domain Survey}
\shortauthors{Wasleske and Baldassare}
\graphicspath{{./}{./}}

\begin{document}

\title{X-ray Emission of Ultraviolet Variable Active Galactic Nuclei Candidates}

\author[0000-0003-3986-9427]{Erik J. Wasleske}
\affiliation{Department of Physics and Astronomy, Washington State University, Pullman, WA 99163, USA}

\author[0000-0003-4703-7276]{Vivienne F. Baldassare}
\affiliation{Department of Physics and Astronomy, Washington State University, Pullman, WA 99163, USA}

\begin{abstract}

We analyze the X-ray properties for a sample of 23 high probability AGN candidates with ultraviolet variability identified in \cite{wasleske2022}. Using data from the Chandra X-ray Observatory and the XMM-Newton Observatory, we find 11/23 nuclei are X-ray detected. We use SED modeling to compute star formation rates and show that the X-ray luminosities are typically in excess of the X-ray emission expected from star formation by at least an order of magnitude. Interestingly, this sample shows a diversity of optical spectroscopic properties. We explore possible reasons for why some objects lack optical spectroscopic signatures of black hole activity while still being UV variable and X-ray bright. We find that host galaxy stellar emission and obscuration from gas and dust are all potential factors. We study where this sample falls on relationships such as $\alpha_{\rm OX}-L_{2500}$ and $L_{X}-L_{IR}$ and find that some of the sample falls outside the typical scatter for these relations, indicating they differ from the standard quasar population. With the diversity of optical spectroscopic signatures and varying impacts of dust and stellar emissions on our sample, these results emphasizes the strength of variability in selecting the most complete set of AGN, regardless of other host galaxy properties.

\end{abstract}


\section{Introduction} \label{sec:intro}

It has been well established that black holes are ubiquitous in the centers of massive galaxies \citep{magorrian1998}. Studies in the growth and evolution of supermassive black holes (SMBHs) have lead to open questions about the occupation and active fractions of black holes (BHs). A complete census of active galactic nuclei (AGN) is important for constraining the occupation fraction and for studying BH growth throughout time. However, this is difficult to obtain due to selection biases in each wavelength regime across the electromagnetic spectrum, as well as observational limitations. Gas, dust, star formation, and host galaxy dilution each have varying effects on masking the activity of the central black hole. This paper is focused on the power of variability selection of AGN for taking a complete AGN census.

The effects of obscuration can be seen across the spectrum of AGN radiation. Absorbing dust emits in the infrared, and can mask ultraviolet to infrared emission while, alongside gas, attenuating X-rays \citep{hickox2018}. The lack of luminous X-ray emission in active galaxies that were bright at other wavelengths, such as the infrared, has been shown to be an indicator of heavy obscuration \citep{donley2012,stern2015,lambrides2020, carroll2021}. 

The infrared regime has been explored using photometry in bands spanning 1 - 1000~$\mu$m. Galaxy dust, star formation, and AGN can all dominate the infrared regime. This dust can be heated by either AGN radiation or by star formation, making it difficult differentiate between the two mechanisms by the galaxies' infrared colors \citep{landt2008,kirkpatrick2015, hainline2016}. \cite{pfeifle2022} derived expressions to relate column density to infrared line ratios. They find that obscured AGN have redder mid-infrared colors than unobscured systems. Additionally, they found a deficiency of X-ray emission and redder colors for those objects whose host galaxies dominate the 12$\mu$m emission.

Optical spectroscopic studies have been done to collect large population of active galaxies within the local universe \citep{ho1997,reines2013}. These selection techniques use line ratios and broad emission thresholds. The broad line emission can be obscured by torus. Studies have show that the majority of AGN lack broad line emission, ie. are type II AGN \citep{villarroel2014}. The narrow line emission can be effected by dilution from star formation within the host galaxy.

These issues are ever so prominent in identifying active dwarf galaxies ($M_* < 1\times10^{10} M_\odot$) whose broad emission lines are fainter and less broad can often be confused with the signatures of supernovae \citep{filippenko1997,pritchard2012,baldassare2016}. Active low-mass galaxies serve as observation tools to constrain the seeding mechanisms of these BH in the early universe \citep{volonteri2008,volonteri2010,natarajan2014}. Populations of active dwarf galaxies have been established using spectroscopic broad and narrow lines \citep{greene2004,dong2012,reines2013,moran2014,molina2021}, variability \citep{baldassare2018,baldassare2020}, X-ray \citep{greene2007,desroches2009,lemons2015,plotkin2016} and radio emission \citep{reines2020}. These multi-wavelength views of active galaxies enables the study of different components of the AGN, but each wavelength regime comes with its own issues and biases. Between decoupling other astrophysical processes from black hole activity (ie. star-formation, illuminated gas and dust) and other technical limitations there is a significant barrier to collecting a complete and accurate sample of AGN. Volume limits create a specific redshift range we can search for active galaxies, with more luminous, less elusive quasars being the only type we are able to systematically search for at the deepest redshifts.

Though X-rays are not immune from obscuration effects, X-ray observations have been well utilized to identify and study the properties of active galaxies (ie. see \cite{brandt2005} and references therein). Sufficiently bright X-ray emission is often used to confirm the presence of AGN in systems lacking other AGN indicators (e.g., \citealt{birchall2020, agostino2022,messick2023} (submitted)). 

\cite{agostino2022} studied local type 2 AGN selected by X-ray emission who lacked apparent optical emission lines. Even with ~25\% of their sample spectra was absorption-line dominated, they found measurable [\ion{O}{3}] emission in their total sample. [\ion{O}{3}] emits far from the torus of the AGN and can serve as an indicator of AGN power. They found a correlation with the specific star formation rate (SFR) and the [\ion{O}{3}] luminosity of these X-ray AGN. \cite{agostino2022} claims that less-luminous [\ion{O}{3}] galaxies contribute to the scatter of their [\ion{O}{3}] to x-ray luminosity relation. This scatter is proposed to be from gas content differences of the narrow line region. They are distinguished only by existing at the low-luminous ($\text{log} \: L_X <42$) end of the broad X-ray to [\ion{O}{3}] uni-modal relation. This separation of less-luminous to luminous [\ion{O}{3}] emitters is due to the specific star-formation rate. The amount of available molecular gas within the host galaxies drives certain processes, as star-formation is dependent on gas, with the photoionization of this gas from the AGN creating its spectroscopic signatures.

\cite{lambrides2020} also investigated this broad X-ray to [\ion{O}{3}] relationship for a population of obscured AGN, finding that the type 2 AGN are below the relation of \cite{yan2011} set for unambiguous AGN. These objects have low observed X-ray emission per their amount of [\ion{O}{3}] emission due not from any intrinsic factors, but from heavy obscuration. Comparison of AGN candidates with varying spectroscopic classes can make indications on the molecular gas contained in the host galaxy. 

Correlations of the X-ray to the mid-infrared (MIR) show a near linear relation for nearby AGN with low-luminosities,  whose dispersions are attributed to the geometry of the dust, star-formation and variability of the AGN \citep{lutz2004,gandhi2009}. This relationship has been use to convert between X-ray and MIR luminosities regardless of the nature of the AGN, as both uncontaminated MIR and X-ray emission are a proxy for the AGN's intrinsic power \citep{asmus2015}. 

\cite{wasleske2022} identified 48 AGN candidates by their high variability in the near ultraviolet (NUV) band. They selected these candidates from ~2000 galaxies from the NASA Sloan Atlas that lay within the sky covered by the GALEX Time-Domain Survey \citep{gezari2013}. The 48 AGN candidates had a variety of optical spectroscopic properties, with some lacking optical spectroscopic AGN signatures entirely. In this paper, we search for nuclear X-ray emission from these UV-variable sources to search for additional evidence of the presence of AGN. Furthermore, we compare their X-ray luminosities to various empirical relations to search for differences between spectral classes. This analysis is compared to that analysis work of \cite{wasleske2022}, which used optical spectroscopy, mid-infrared colors, and low-resolution spectral energy distributions. In the analysis of the optical spectra of these variable galaxies, \cite{wasleske2022} found examples of their variable population having signatures of AGN, star formation, composites, and even absorption-line dominated systems.

This study compares the variable AGN population to known X-ray relations of AGN emission to investigate the underlying physics of their emission signatures. Our goal is to measure X-ray emission from variable AGN candidates and use the multi-wavelength galaxy properties to determine whether dust, obscuration, and star-formation impact the optical spectral signatures.

This work is structured as follows: In Section \ref{sec:data}, we discuss our sample and the data available in the \textit{Chandra X-ray Observatory} (CXO) and \textit{XMM-Newton} (XMM) archives. In Section \ref{sec:analysis}, we discuss the data reduction and the construction of spectral energy distribution (SED) models. In Section \ref{sec:results}, we report our results from the X-ray emission and SED modeling. In Section \ref{sec:discussion}, we discuss the implications of these results in relationship to literary observational relations, associating optical, IR, and X-ray.

\section{Data} \label{sec:data}

\subsection{Sample Selection} \label{subsec: sample selection}
Our sample is comprised of all objects from \cite{wasleske2022} that had available X-ray data within the CXO or XMM archives. \cite{wasleske2022} identified 48 high-probability AGN candidates from their near-ultraviolet variability in the GALEX Time-Domain Survey (TDS; \citealt{gezari2013}). The parent population for this variability study was taken from the NASA Sloan Atlas (NSA), which combines imaging products from SDSS and GALEX to construct a catalog of roughly 641,500 galaxies with redshifts and other derived quantities \citep{blanton2009,maller2009,zhu2010}. There were $\sim2,000$ galaxies in the region covered by GALEX TDS which were analyzed for their NUV variability. The 48 high-probability candidates were selected for significant variability in their light curve over the 3 year baseline of the TDS. The pointing of the TDS covered six of the Pan-STARRS1 Medium Deep Survey fields, four of which are in the northern sky. The NUV filter covers a range from 1750-2800\AA.  

Of the 48 AGN candidates, 23 unique objects had coverage with either CXO or XMM Archive. The description of these data products is given below.

\subsection{Chandra X-ray Observatory} \label{subsec: CXO data}

We searched for 48 AGN candidates from \cite{wasleske2022} in the CXO Archive. To establish whether objects were observed with Chandra, we searched for each object in the Chandra Archive with the default search radius of 10 arc-minutes. We then examined each observation to determine whether the object was contained within the field. We found a total of 13 objects covered by Chandra ACIS observations, with some objects having repeat observations. Reduction of these observation products is described in \ref{subsec: ciao reduction}. 
NSA 64129 is the only galaxy that was not analyzed. It is 9.07 arc-minutes from the nearest observation's center, and inspection of that observation's image shows that the galaxy straddles the edge of the CCD, thus we removed it from the CXO sample. NSA 64129 however lays within the XMM source catalog, as discussed in Section \ref{subsec: XMM data}. 
Cross-check of the \textit{Chandra} Source Catalog, CSC 2.0, \citep{evans2020_AAS_abstract_CSC2.0} was done to ensure that the appropriate data products were collected from the archive for the remaining 12 objects.

Our sample of 12 objects having CXO observations have an array of optical spectroscopic signatures that were analyzed in \cite{wasleske2022}. From the BPT diagram \citep{baldwin1981}, two objects are classified as AGN, one as starburst, one as a Composite galaxy. Three of the objects are dominated by absorption features while the rest either have no available SDSS spectrum or spectrum too noisy to preform the fitting routine. 

\subsection{XMM-Newton Observatory} \label{subsec: XMM data}

We searched the XMM-Newton Serendipitous source catalog \citep{webb2020} for matches to our 48 AGN candidates from \cite{wasleske2022}. Beyond collected detections, we searched for upper limits of the remaining galaxies within XMM's coverage.

The 4XMM-DR9 catalog is based off a set of 11,204 EPIC observations that contain a detection, spanning 19 years. Data processing for this catalog was based on Science Analysis Software (SAS) version 18 to produce calibrated event lists for each observation. The process of converting the raw observation data event files from the EPIC instruments into event lists by the pipeline is the same as described in \cite{watson2009}. The 4XMM-DR9 updates to the 2XMM \citep{watson2009} and 3XMM \citep{rosen2016} catalogs are the inclusion of source spectra, light curves, source detection, and event corrections using Timing mode and pn small window data. This catalog contains ~550,000 unique sources, building upon earlier iterations with an improved source detection and provides an ideal easy access to data products from the \textit{XMM Newton} archive. \cite{webb2020} discusses the properties of the catalog, finding a Gaussian distribution of fluxes for all sources detected. The minimum flux value for the 0.2 - 12 keV broadband is $1.44 \times 10^{-19}$erg/cm$^2$/s, for source 4XMM J174538.0-285950. 

This catalog contains fluxes in 9 X-ray bands for each object that span different pieces of $0.2 -12 \rm{keV}$ energy range. Analysis of the selected sources is described in Section \ref{subsec: source catalog}. Of the 48 variable AGN candidates, seven were found in the source catalog. Selection and analysis of these sources is discussed in Section \ref{subsec: source catalog}.

Using the FLIX server\footnote{http://flix.irap.omp.eu/}, we compute X-ray upper limits for an additional 14 of the AGN candidates from \cite{wasleske2022}, given in Table \ref{tab:lum table}. These 14 objects were within fields observed by XMM, but the separation between the nearest detected source within the catalog to each of these candidates was within the range of 50 - 207 arcsec. Seven of these objects were also found in the CXO Archive.  We compute upper limits using the pn band 8 (0.2 -12.0 keV) flux from the observation with the minimum Axis Offset value for each of these 14 candidates. 

\subsection{Ancillary Data} \label{subsec: SED data}

Photometry was collected from surveys across the electromagnetic spectrum to be used for input values of the SED models. Along with the photometric X-ray values pulled from our analysis, we use the median GALEX photometric values from \cite{wasleske2022} for each variable galaxy, data from SDSS DR16 \citep{ahumada2020}, the United Kingdom Infrared Telescope (UKIRT; \cite{lawrence2007}), Wide-field Infrared Survey Explorer (WISE; \cite{wright2010}), unblurred and unofficial coadds of the WISE imaging (unWISE; \cite{lang2014}), and the Two Micron All-Sky Survey (2MASS; \cite{skrutskie2006}). 

The compilation of these values serve as the input of the SED analysis using X-CIGALE V2022.0 \citep{yang2022}, discussed in Section \ref{subsec: sed models}.

\input{table1}

\section{Analysis} \label{sec:analysis}
We present an examination of the reduction and compilation of X-ray data to measure broadband photometric luminosities. We also discuss the process and the choices made for the SED modeling.

\subsection{CIAO Reduction} \label{subsec: ciao reduction}

The Chandra Interactive Analysis of Observations (CIAO, version 4.14 \footnote{see https://cxc.cfa.harvard.edu/ciao/ for user guides of this software package}) data reduction software was used to reduce and analyze the observations. For each observation file, we first reprocessed the observation files using the $\texttt{chandra\_repro}$ script to apply the latest calibrations, which creates a new level 2 event file and a bad pixel file from the background cleaning of the original event file. Preliminary sources within the level 2 event file are then identified using CIAO $\texttt{WAVDETECT}$ function. $\texttt{WAVDETECT}$ does this through correlating potential pixel sources to Ricker wavelet functions, removing highly correlated values as assumed sources. After this is completed, we filter the event file to the 0.5-10.0 keV range. 

We used CIAO's $\texttt{SRCFLUX}$ software to carry out the aperture photometry. $\texttt{SRCFLUX}$ is a software that calculates net count rates and fluxes with uncertainties while accounting for contributions of the point spread function to both source and background apertures. We use a $2.0"$ source aperture in unison with a source-less background annulus of size $20.00"$ for its inner and $30.00"$ for its outer radius. Our source aperture is set by the spatial resolution of the CXO. 

The spectral model we used within the $\texttt{SRCFLUX}$\footnote{$\texttt{SRCFLUX}$  references its models from Xspec, see https://heasarc.gsfc.nasa.gov/xanadu/xspec/manual for user manual} function for all CXO objects is $\texttt{xsphabs.abs1 * xspowerlaw.pow1}$ where $\texttt{xspowerlaw}$ is a simple photon power law where we defined its index as $\Gamma = 1.8$. This model accounts only for galactic absorption, not the host galaxy of each for these sources. We set the absorption parameter to $\texttt{abs1.nH=\%GAL\%}$ so that National Radio Astronomy Observatory values for galactic column density $n_H$ are collected.
 
Assuming this absorbed power law spectrum we then collect count and flux estimates for broad (0.5 - 7.0 keV), soft (0.5 - 1.2), medium (1.2 - 2.0 keV), and custom 2.0 - 10.0 keV bands. Once these estimated values are collected, we follow the method of \cite{baldassare2017} to measure the intrinsic absorption. Using the Portable Interactive Multi-Mission Simulator (PIMMS) toolkit\footnote{https://heasarc.gsfc.nasa.gov/docs/software/tools/pimms.html}, we started with our custom 2.0 - 10.0 keV band's count rate to calculate the expected count rate for the 0.5 - 2.0 keV band, assuming a photon index of $\Gamma = 1.8$. If this expected value was higher than the observed rate from the CIAO reduction, an intrinsic $n_H$ was calculated. This was the case for four of the objects detected in CXO, with the intrinsic $n_H$ value given in Table \ref{tab:lum table}.

From these fluxes, we calculate the luminosities using the distances given in the NSA.

\begin{figure*}[h]
    \centering
    \subfigure{\includegraphics[scale=0.42]{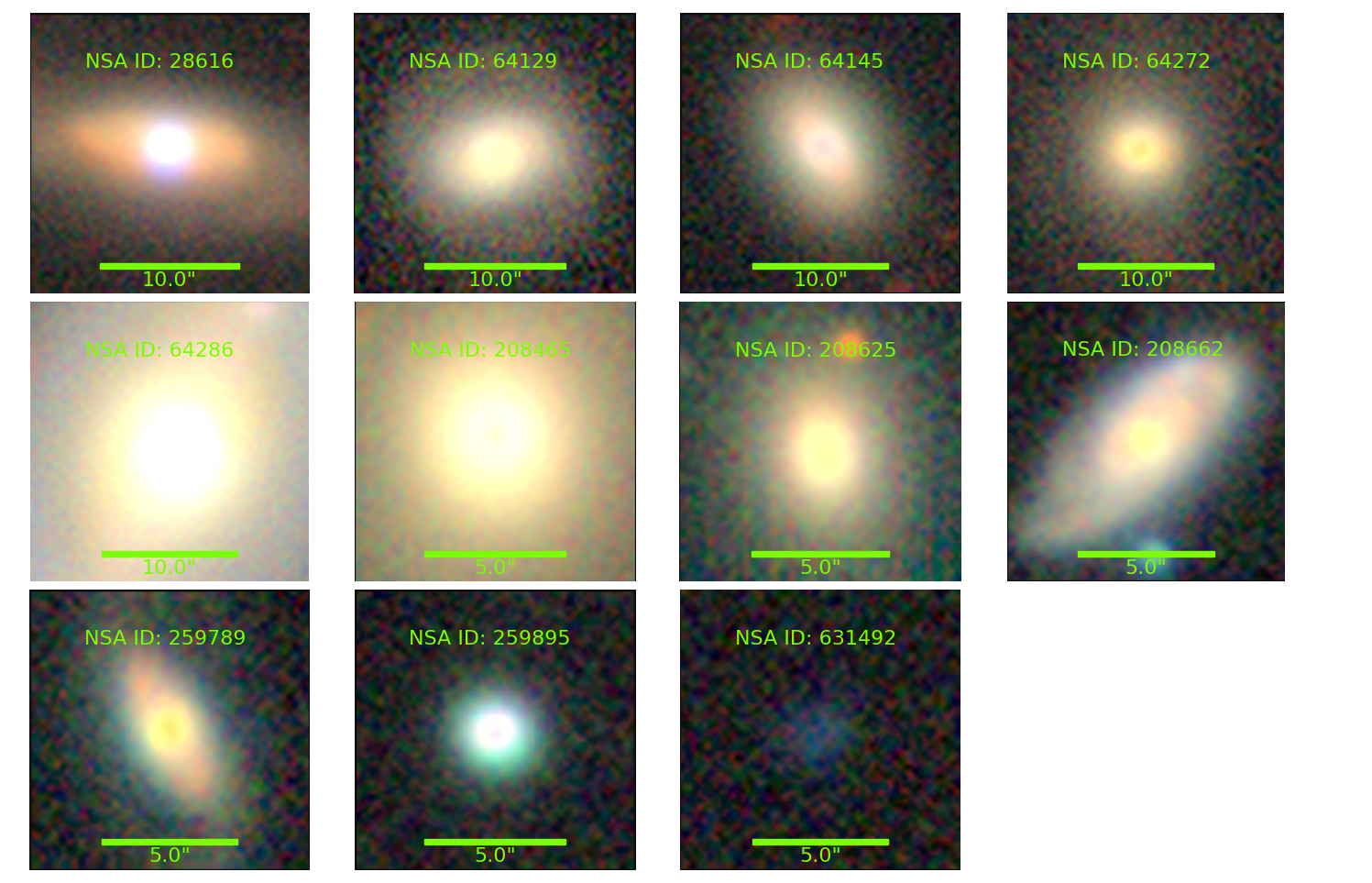}}
    \hspace{0.0em}
        \subfigure{\includegraphics[scale=0.42]{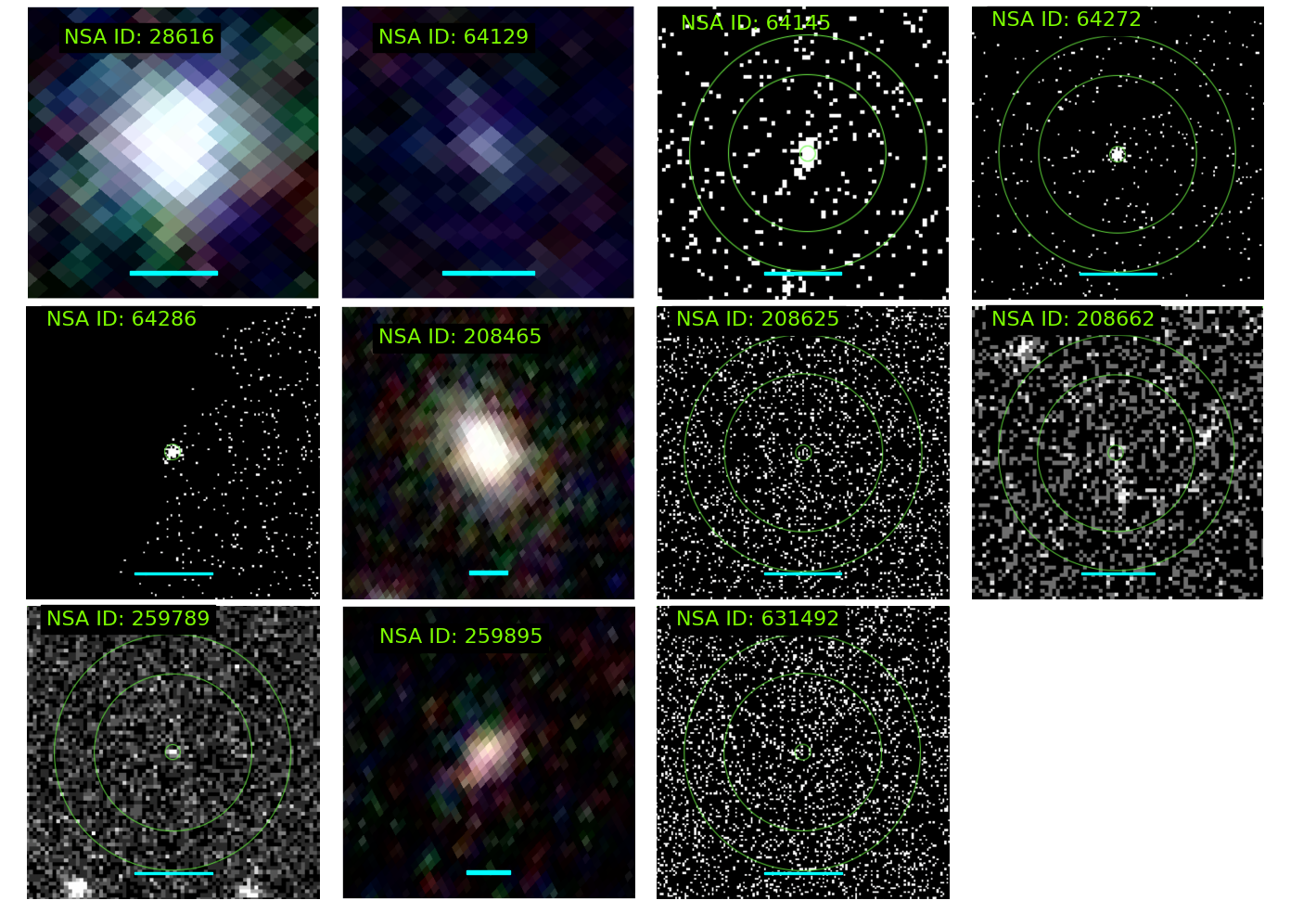}}
    \caption{\emph{Top:} Legacy Survey images of the galaxies from our sample with X-ray detections in either CXO and XMM. \emph{Bottom:}  X-ray images for each galaxy with an X-ray detection. The teal scale-bar is 20" on each panel. For objects detected with XMM (i.e. NSA 28616, 64129, 208465, and 259895), the XMM EPIC-RGB cutout is shown \footnote{collected from http://xmm-catalog.irap.omp.eu/}. For objects with detected in CXO, the circular source aperture and annulus background aperture from the CIAO data reduction are given in green. For those detected in both CXO and XMM, the CXO image is displayed. }  
    \label{fig:Legacy&CXO_Images}
\end{figure*}

The bottom panel of Figure \ref{fig:Legacy&CXO_Images} shows the source and background apertures over each event file used in the data reduction. For NSA 64286, as the object was collected near the edge of the ACIS chip, the background annulus apertures was centered at \text{RA} = 150.500\textdegree, \text{DEC} = 3.077\textdegree, 2.1' away from the objects position and the source aperture. Analogous optical Legacy Imaging cutouts are given in the top panel of Figure \ref{fig:Legacy&CXO_Images}. 

\subsection{XMM Source catalog objects} \label{subsec: source catalog}

\begin{figure*}[h]
  \centering
    \includegraphics[width=\textwidth]{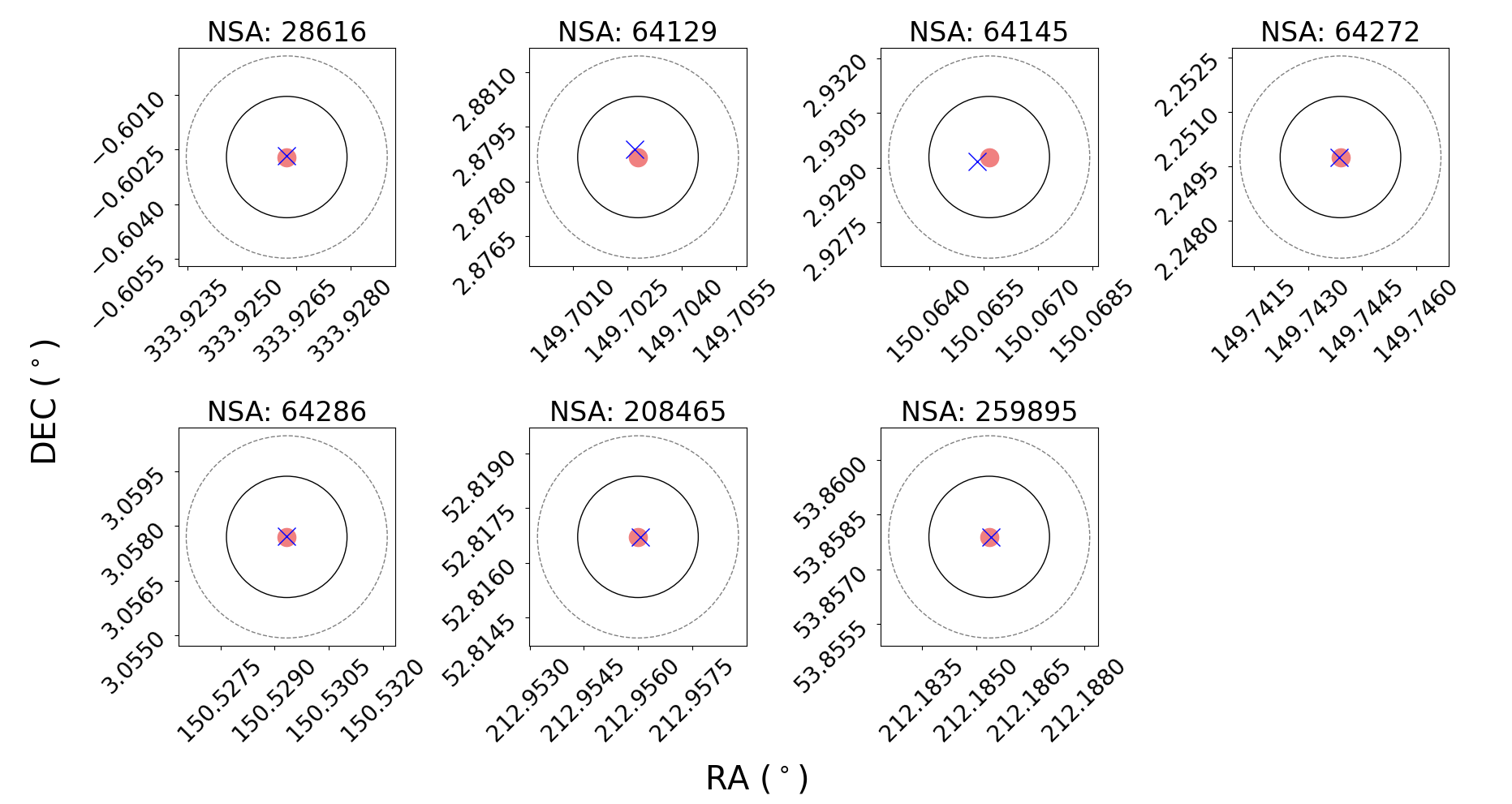}
  \caption{X-ray source and object positions for sources found in the XMM-Newton catalog. Faint red circles show the object position, while the blue "X" represents the X-ray source position. The solid black circle represents the 6" FWHM point-spread function of XMM. The 10" threshold circle is given in dashed grey.
  }
  \label{fig: XMM_source_obj_pointing}
\end{figure*}

Sources from the \textit{XMM-Newton Serendipitous Source catalog} were cross matched to the AGN candidates via closest 2-D sky separation. This list was then filtered down to more probable comparison if the sources were within a 10" radius of the objects position. This separation threshold filtered the list down to seven probable sources, three of which overlap with our CXO sample. Figure \ref{fig: XMM_source_obj_pointing} shows the positions of the source compared to the position given in the NSA catalog. The 6" resolution of XMM instrument is given by the black circles on each panel of Figure \ref{fig: XMM_source_obj_pointing}. This demonstrates the accuracy of the 2-D position cross matching used to select these sources. We collect the $0.2 -12~\rm{keV}$ fluxes for these seven sources.

Once these sources were identified, we used the Spectrum Fitting tool\footnote{ see 'Fit Spectrum' option of http://xmm-catalog.irap.omp.eu/ for Source and respective Detection IDs.} for these source to fit the broadband 0.2-12.0 keV spectrum. This fitting gave a total absorption estimate for the spectrum. Since we know the Galactic absorption for each position, we can then compute intrinsic absorption of each object. These values are given for the XMM source detections in Table \ref{tab:lum table}. 

The $0.5 - 7.0~\rm{keV}$ broadband luminosity for these X-ray analogous sources and the FLIX upper limits are given in Table \ref{tab:lum table}. The PIMMS toolkit was used to convert the broadband luminosity $0.2 -12~\rm{keV}$ to the broad $0.5 - 7.0 ~\rm{keV}$ CXO band for ease of comparison. 

\subsection{Spectral Energy Distribution Modeling} \label{subsec: sed models}

Using \texttt{X-CIGALE V2022.0} \citep{yang2022}, we fit spectral energy distribution (SED) models to estimate emission mechanisms for all 23 objects. The photometric values, including those for the GALEX NUV band, are equivalent to those used in the \cite{wasleske2022} SED decomposition, with the addition of the X-ray values collected above. X-ray boxcars for the ranges of $0.5-7.0 \rm{keV}$ and $0.2-12.0 \rm{keV}$ were used for CXO and XMM sources respectively. We followed the SED-fitting methodology of \cite{burke2021} to set the specific models and parameter ranges for our SED models. \cite{burke2021} SED analysis focused on modeling galaxies whose SEDs were not dominated by the central AGN, which is well-aligned with the sample and science goals presented here. We detail this work below.

To model the AGN emission, we use the $\texttt{SKIRTOR}$ \citep{stalevski2012,stalevski2016} model, for its two-phase clumpy model of the radiative transfer of the AGN's dusty torus. We let the AGN fraction vary from 0.0 to 0.9, and the polar extinction $E(B-V)$ maximum value be $0.175$. Viewing angle was let to vary between 10\textdegree and 60\textdegree. 

The star formation history was done using a delayed exponential model. This model used parameters of \textit{e}-folding times ranging from 100 - 5000 Myr and age of main stellar population of 1000, 2000, or 5000 Myr. We used the initial stellar mass function from \cite{chabrier2003}, the stellar population models of \cite{bruzual2003}, and nebular emission model of \cite{inoue2011}. Dust emission was modeled by \cite{draine2014}, with extinction reddening from \cite{calzetti2000} template. Example SED models are given in the Appendix.

\section{Results} \label{sec:results}
We present the results of the data reduction processes stated above. 

\subsection{X-ray Properties} \label{subsec: broadband emission}

\begin{figure*}
  \centering
    \includegraphics[width=\textwidth]{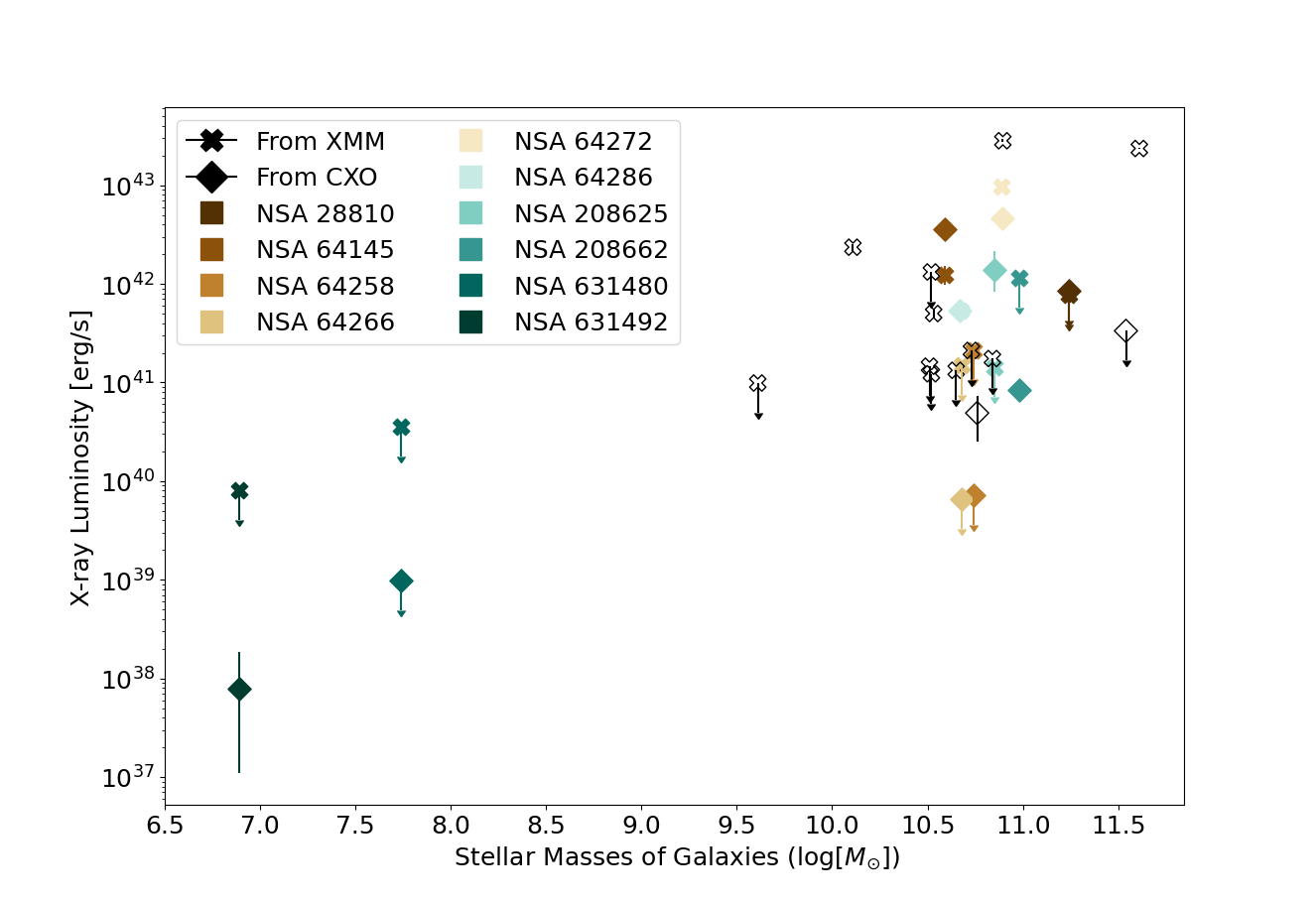}
  \caption{Galaxy mass versus X-ray luminosity for each source and upper limit. Diamonds represent those values collected from CXO, with 'X's representing XMM. We color coordinated points from XMM and CXO to which represent one singular object. Unfilled points represent those objects with values taken from only that one survey.}
  \label{fig: mass - lum plot}
\end{figure*}

Broad band 0.5-7.0 keV X-ray luminosities are given in Table \ref{tab:lum table} for the 23 objects covered by CXO and/or XMM. X-ray luminosities range from $L_{0.5-7.0 \rm{keV}} = 7.8 \times 10^{37}$ to $2.8 \times 10^{43}$ $\rm{erg~s^{-1}}$. Of the 23 galaxies with X-ray coverage in either CXO or XMM, 11 are detected in X-ray. Figure \ref{fig: mass - lum plot} plots the relationship of stellar mass to $0.5 - 7.0~\rm{keV}$ luminosity for both XMM and CXO sources. 

Through the CXO data reduction for 12 objects, we found X-ray luminosities ranging from $L_{0.5-7.0 \rm{keV}} = 7.8 \times 10^{37}$ to $4.6 \times 10^{42}$ $\rm{erg~s^{-1}}$. There were two dwarf galaxies with CXO coverage - NSA 631492 and 631480. The stellar mass range of objects from the CXO Archive is $\log ( \text{M}_{*} / \text{M}_\odot ) = 6.89$ to $11.54$. 

Upper limits were found for NSA 28810, 64258, 64266, 613397, and 631480 within CXO observations. Their values range from $9.8 \times 10^{38}$ $\rm{erg~s^{-1}}$ for NSA 631480 to $8.6 \times 10^{41}$ $\rm{erg~s^{-1}}$ for NSA 28810.

The X-ray luminosities for the XMM source catalog objects are generally higher than for those observed with CXO. The X-ray luminosities range is $L_{0.5-7.0 \rm{keV}} = 5.3 \times 10^{41}$ to $2.8 \times 10^{43}$ $\rm{erg~s^{-1}}$. The catalog sources are all massive galaxies, with stellar mass range of $\log ( \text{M}_{*} / \text{M}_\odot ) = 10.53$ to $11.61$. 

Upper limits from XMM observations were found for NSA 28810, 64258, 64266, 205160, 208625, 208662, 208702, 259478, 259880, 259919, 260221, 260241, 631480, and 631492. Their values range from $8.0 \times 10^{39}$ to $1.3 \times 10^{42}$ $\rm{erg~s^{-1}}$. We note these provide loose constraints on the luminosity of the objects as these limits are derived from flux upper limits of the observations themselves.

NSA 631492 has the lowest luminosity of our whole X-ray population, at $L_{0.5-7.0 \rm{keV}} = 7.8 \times 10^{37}$ $\rm{erg~s^{-1}}$. It is also worth noting that NSA 631492 is the least massive object in this study, at $\log ( \text{M}_{*} / \text{M}_\odot ) = 6.89$ and it lacks an optical spectrum from SDSS. In contrast, the most massive galaxy, NSA 208465, with a stellar mass of $\log ( \text{M}_{*} / \text{M}_\odot ) = 11.61$, is the second most luminous within our population. Its optical spectrum is dominated by absorption lines.

We find that three of the seven X-ray sources within CXO are moderately obscured, with intrinsic $n_H$ on the order of $10^{22} \text{cm}^{-2}$. Comparatively, all the XMM sources are below this threshold of obscuration, pointing to the bias of the XMM source catalog towards bright AGN.

Figure \ref{fig: mass - lum plot} shows the stellar masses versus the $L_{0.5-7.0 \rm{keV}}$ for each target. We observe a general trend that an increase in host stellar mass leads to an increase in X-ray luminosity. Objects with both CXO and XMM luminosities are marked on the plot. These cases are discussed further in Section \ref{subsec: cross listed objects}. 

\subsection{Objects detected with CXO and XMM} \label{subsec: cross listed objects}

\begin{figure*}[h]
    \centering
    \subfigure{\includegraphics[scale=0.45]{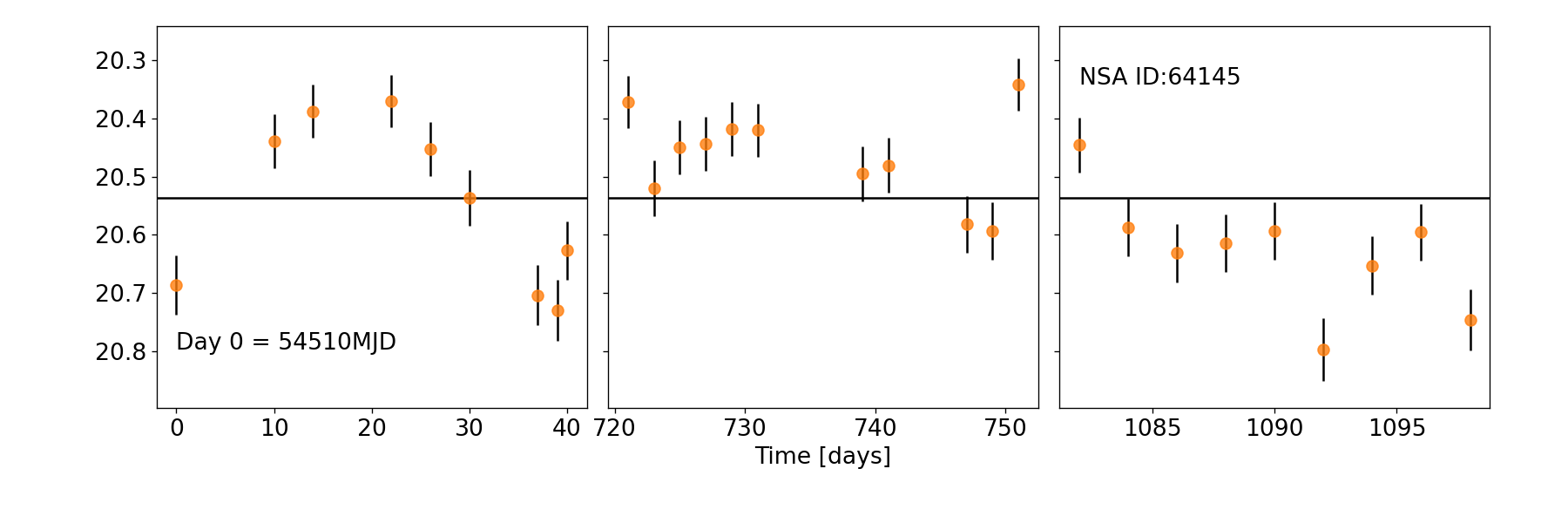}}
    \hspace{1em}
    \subfigure{\includegraphics[scale=0.45]{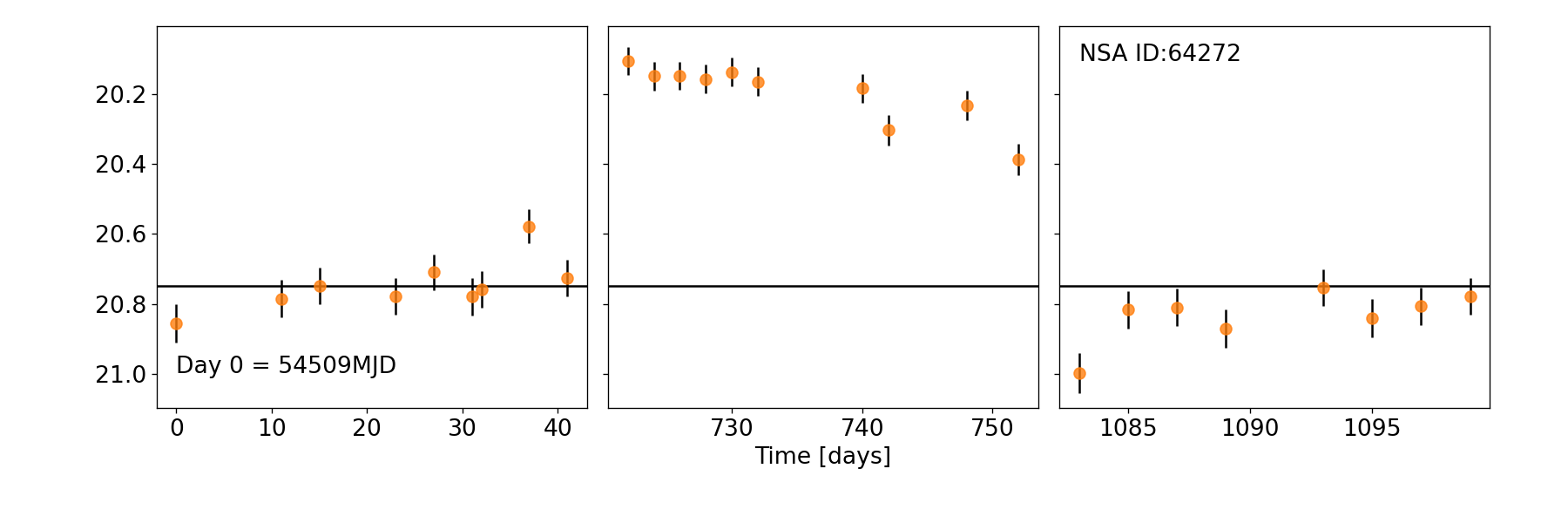}}
    \hspace{1em}
    \subfigure{\includegraphics[scale=0.45]{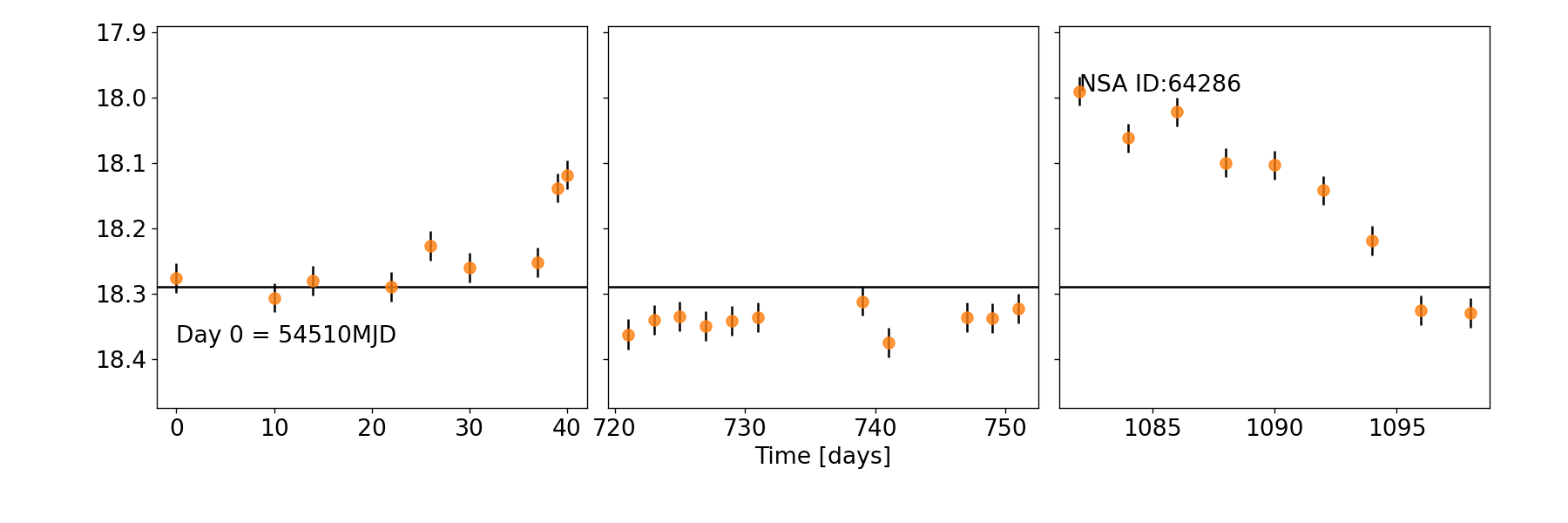}}
    
    \caption{\emph{Top:} GALEX NUV light curve of NSA 64145 from \cite{wasleske2022}. \emph{Middle:} GALEX NUV light curve of NSA 64272. \emph{Bottom:} GALEX NUV light curve of NSA 64286. Each of these light curves so significant ($>0.2 \textit{mag}$) variability across roughly three years. NSA 64272 and 64286 show significant change in mean magnitude in a one of their observation windows, for a duration of about a month. Thus, we investigate the possibility of a TDE in Section \ref{subsec: cross listed objects}. }
    
    \label{fig:light curves of 64145 64272 64286}
\end{figure*}

As seen in Table \ref{tab:lum table}, NSA 64145, NSA 64272, and NSA 64286 have both bright X-ray emission within archival CXO data and an associated source within the XMM Source Catalog. Thus, we can evaluate whether these sources have persistent or variable/transient X-ray emission. We present the GALEX near-UV light curves for these three objects from \cite{wasleske2022} in Figure \ref{fig:light curves of 64145 64272 64286}.

We find consistent and persistent broadband X-ray emission for NSA 64145, NSA 64272, and NSA 64286. For NSA 64145, the CXO observations for CXO were taken in June 2013 and the XMM observations were taken between December 2004 and May 2006. Thus, NSA 64145 demonstrates persistent X-ray emission over near decade-long timescale. NSA 64286 also has consistent X-ray luminosity between its XMM and CXO observations, taken in May 2015 and March 2019 respectively. For NSA 64272, both its CXO and XMM X-ray luminosities are roughly consistent,within a half dex of each other. The XMM source catalog used observations from December 2003, May 2006, and May 2007. The CXO observation for this object was taken in January of 2007. 

We consider the X-ray emission from the three sources above to be most likely due to an AGN as opposed to a variable source such as a tidal disruption event (TDE). TDEs would not be visible in the optical spectrum beyond a time scale of months to a year. Faint and fast TDEs are seen to peak within a month and decline in a shorter time \citep{charalampopoulos2022}. Optical and X-ray emission from TDEs is typically seen over a months-long timescale \citep{rees1988,komossa2008,strubbe2009,wang2012,cao2018} (see \cite{gezari2021} for review and reference therein).

Objects NSA 28810, 64258, 64266, 208662, 631480 and 631492 each have analogous XMM upper limits paired with their CXO detection or upper limit. In these cases, if the object is detected by CXO, the CXO X-ray luminosity is less than the XMM upper limit. 

Interestingly, NSA 208625 has a CXO detection at a higher value than its XMM upper limit. All archival CXO observations of this object were taken between August and October of 2005. XMM observed this field on December 26th, 2015. The UV light curve, constructed with data collected between May 4th, 2009 and March 14th, 2011, shows stochastic variability rather than a burst-like light curve. A possible scenario for this object is a TDE that occurred prior to the CXO observations and faded by the time of the XMM observations. The intermediate UV variability could be from residual accretion following the TDE. Alternately, this object could be a changing state AGN. Further study of object is needed to confirm either scenario.

\section{Discussion} \label{sec:discussion}

We compare X-ray luminosities to different host galaxy properties estimated by the SED modeling. We show that the X-ray luminosities are higher than would be expected based on star formation rates. We compare these objects against known relationships for lumionus quasars, and search for reasons why these UV variable, X-ray bright AGN candidates can lack optical signatures of BH activity. 

\subsection{Origin of X-ray Emission} \label{subsec: origin of x-ray -- SFR vs X-ray}

We can compare X-ray emission to the star formation rates (SFR) as estimated by the SED models as an additional confirmation of the presence of an AGN. \cite{ranalli2003} developed a relation between SFR and hard X-ray luminosity for normal (non-AGN) galaxies. This relation reflects the expected X-ray emission from X-ray binaries and supernova remnants. An excess of X-ray emission for a galaxy's SFR can indicate the presence of an AGN in a galaxy. This relationship was used by \cite{agostino2021} to identify AGN candidates.  

\begin{figure*}
  \centering
  \includegraphics[width=0.8\textwidth]{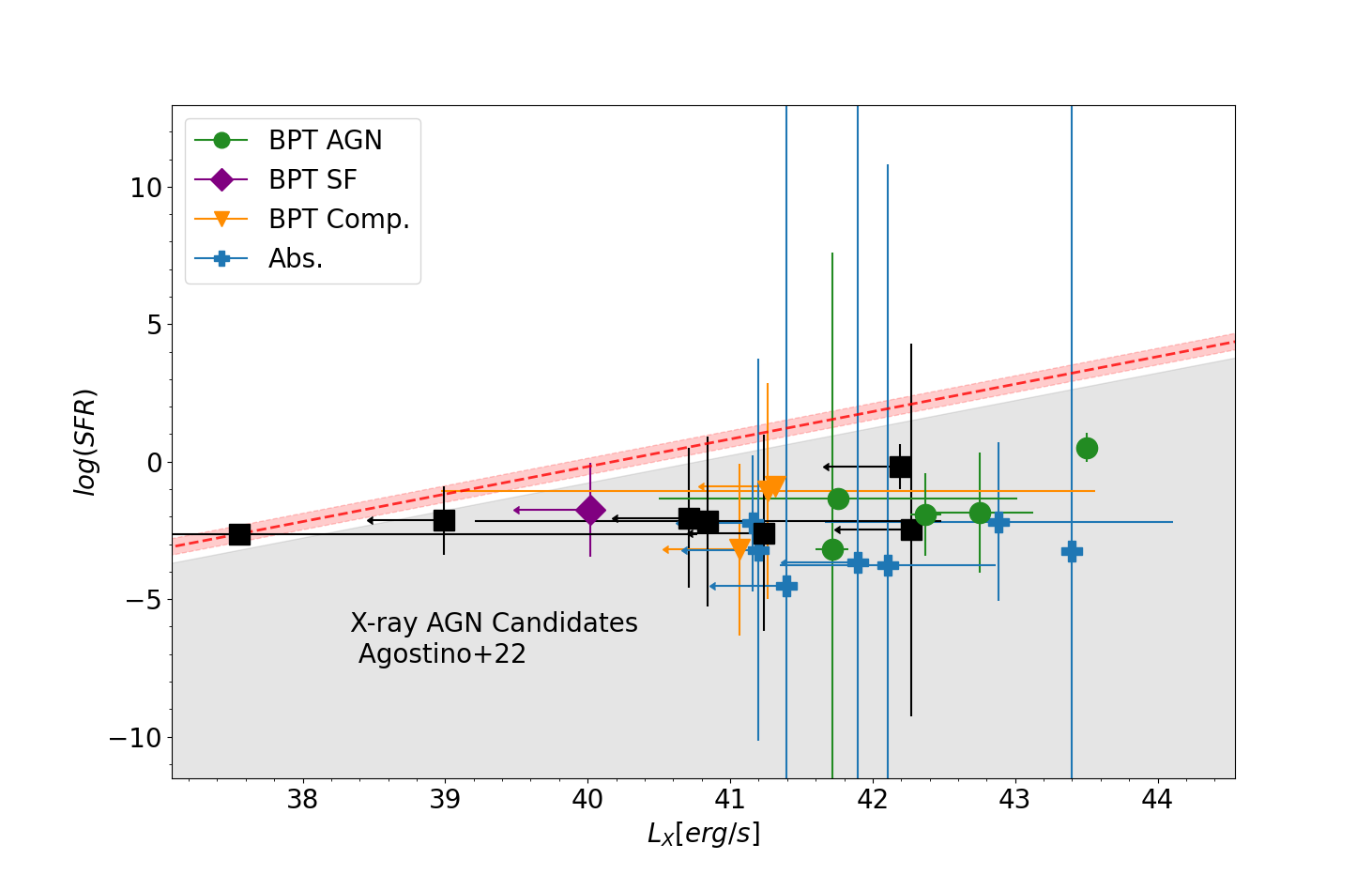}
  \caption{SFR versus 0.5-10 keV X-ray luminosity. The SFRs are the Bayesian SFRs estimated from our SED modeling. The relation between SFR and X-ray luminosity for normal galaxies is defined by \cite{ranalli2003}. \cite{agostino2019} adapted this relation for 0.5-10 keV X-ray luminosities. The dashed red line shows this adaptation, with the shaded red region quoting the intrinsic scatter of $0.3$dex ($1\sigma$) found by \cite{ranalli2003}. The grey shaded region reflects the selection criteria for X-ray AGN from \cite{agostino2021, agostino2022}, where AGN candidates are chosen to have an X-ray excess of $\geq 0.6$dex compared to the \cite{ranalli2003} relation. Green dots, purple diamonds and orange triangles represent BPT AGN, star-forming, or composite classifications on the BPT diagram \citep{baldwin1981} respectively. These values are taken from  \cite{wasleske2022}. Blue crosses represent absorption line dominated spectrum. Black squares represent those galaxies whose either did not have optical spectrum or spectrum that could not be fit.} 
  \label{fig: L_x vs SFR}
\end{figure*}

We plot the estimated SFR from SED modeling versus the $L_{0.5 - 10 \rm{keV}}$ for our sample in Figure \ref{fig: L_x vs SFR}. Almost every galaxy in our sample is shown to have excess X-ray emission compared to the \cite{ranalli2003} relation for non-AGN objects. The notable exception is NSA 631492, which is the lowest mass, lowest X-ray luminosity object in the sample. 

\cite{agostino2022} finds that LINER type galaxies have excess X-ray emission across broad 0.5-10 keV for their SFR compared to normal non-AGN galaxies. They separate their X-ray AGN candidates as those having X-ray luminosities more than 0.6 dex above the X-ray luminosity predicted from their SFR. All but NSA 631492 lay within \cite{agostino2022} criteria for X-ray AGN.

Even the galaxies with SF, composite and absorption line dominated optical spectrum in our sample have X-rays in excess of the SFR predictions. The galaxies with optical spectroscopic AGN signatures have a higher excess of X-ray emission compared to the spectroscopic star-forming and composite galaxies. Our results are similar to those of \cite{agostino2019}, who found that some of their X-ray AGN candidates did fall within the star forming region of the BPT diagram. \cite{agostino2022} also finds that one-quarter of their X-ray AGN sample have apparent pure absorption-dominated spectra. However, it is possible that higher resolution optical spectra (compared to SDSS spectroscopy) or more robust optical spectroscopic modeling would reveal some AGN signatures. 

All of the X-ray detected objects besides NSA 631492 are likely AGN based on having UV variability and X-ray emission in excess of SFR-based expectations. This includes all 5 BPT AGN, 1/3 BPT composite, 3/7 absorption line systems, and 1/7 of the galaxies with with no optical spectroscopy. For the non-detected objects, the upper limits are generally not constraining enough to rule out the possible presence of an AGN. 

These results demonstrate variability selection is as effective as X-ray selection at identifying AGN in objects lacking optical spectroscopic signatures. 


\subsection{Diversity of Optical Spectroscopic Properties} \label{subsec: reasons for spec}

 Through our X-ray analysis, we establish that almost all of the UV-variability selected, X-ray detected objects are bonafide AGN. In this section, we discuss possible explanations for diversity in optical spectroscopic signatures.

\subsubsection{Stellar populations} \label{subsubsec: stellar effects}

\begin{figure*}
    \centering
    \includegraphics[width=0.8\textwidth]{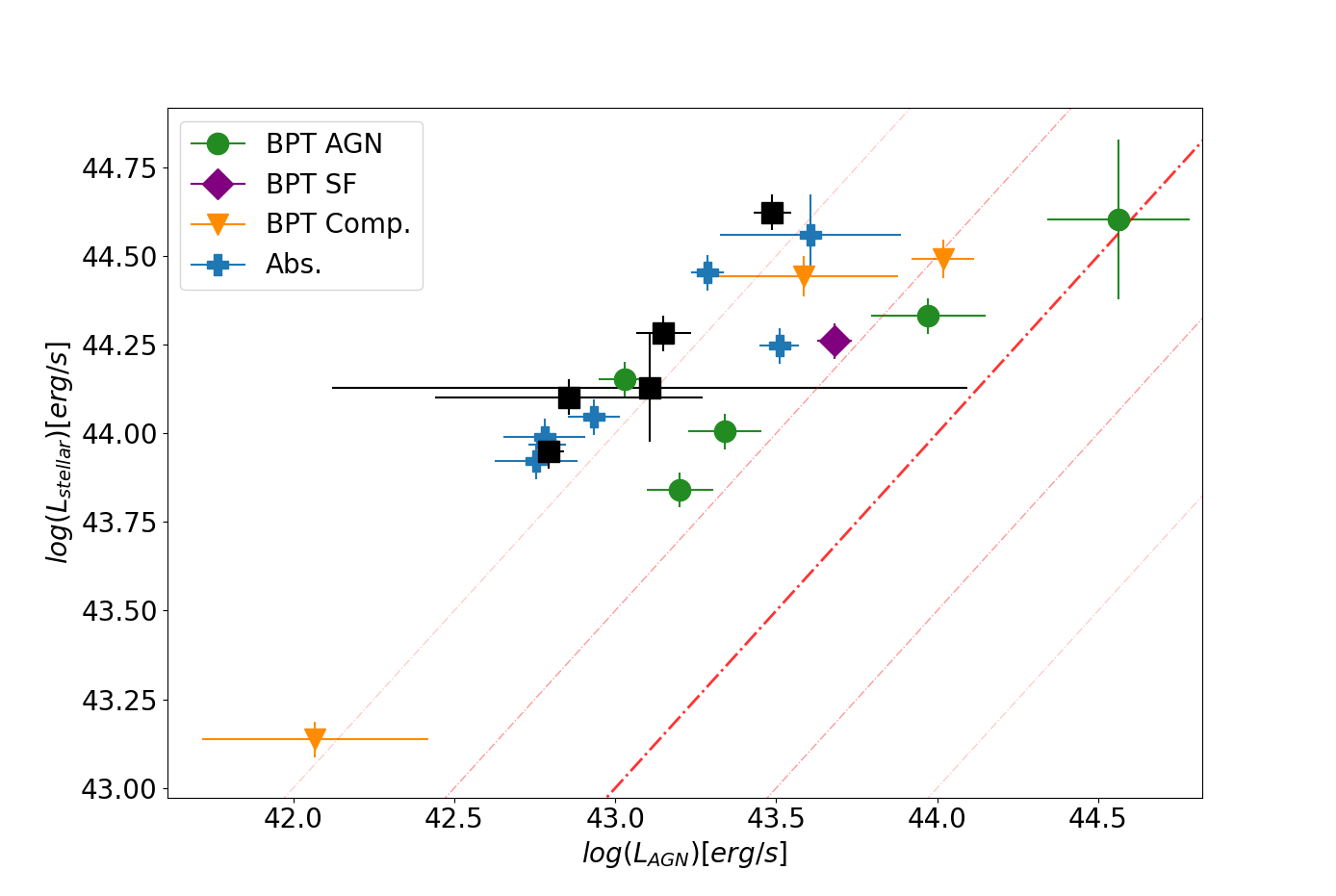}
    \caption{Scatter plot of AGN to stellar luminosity measured from SED modeling. The thick red dash-dot is the unity line. The other faded red curves are 0.5 dex deviations from unity. Points for NSA 631492 and 631480 are left off as they exist in the far left, less luminous regime, compared to the rest of the sample. Marker shapes and colors are the same as those defined for Figure \ref{fig: L_x vs SFR}.}
    \label{fig:stellar_relation}
\end{figure*}

Here, we investigate the AGN power versus the emission from stars, as estimated by our SED modeling. We note that all of the objects in our sample preferred an AGN to be included in the best-fit SED. For galaxies without optical spectroscopic AGN signatures, it is possible that host galaxy stellar light is diluting the AGN emission. Stellar and AGN emission can also both heat dust. In Figure \ref{fig:stellar_relation}, we show the best fit stellar luminosity versus AGN luminosity from SED modeling. For the full sample, the stellar population is brighter than the AGN. However, the stellar population is more dominant for the absorption line-dominated systems than for the AGN or composite objects. This is notable as the X-ray emission from the absorption line systems is comparable to the BPT AGN, suggesting it is not simply a question of AGN power. This suggests that host galaxy stellar populations or heating of dust by stellar emission could be diluting the AGN signatures for these objects.

\subsubsection{Disk - Corona Relation} \label{subsubsec: alpha ox relation}

\begin{figure*}
  \centering
    \includegraphics[width=0.8\textwidth]{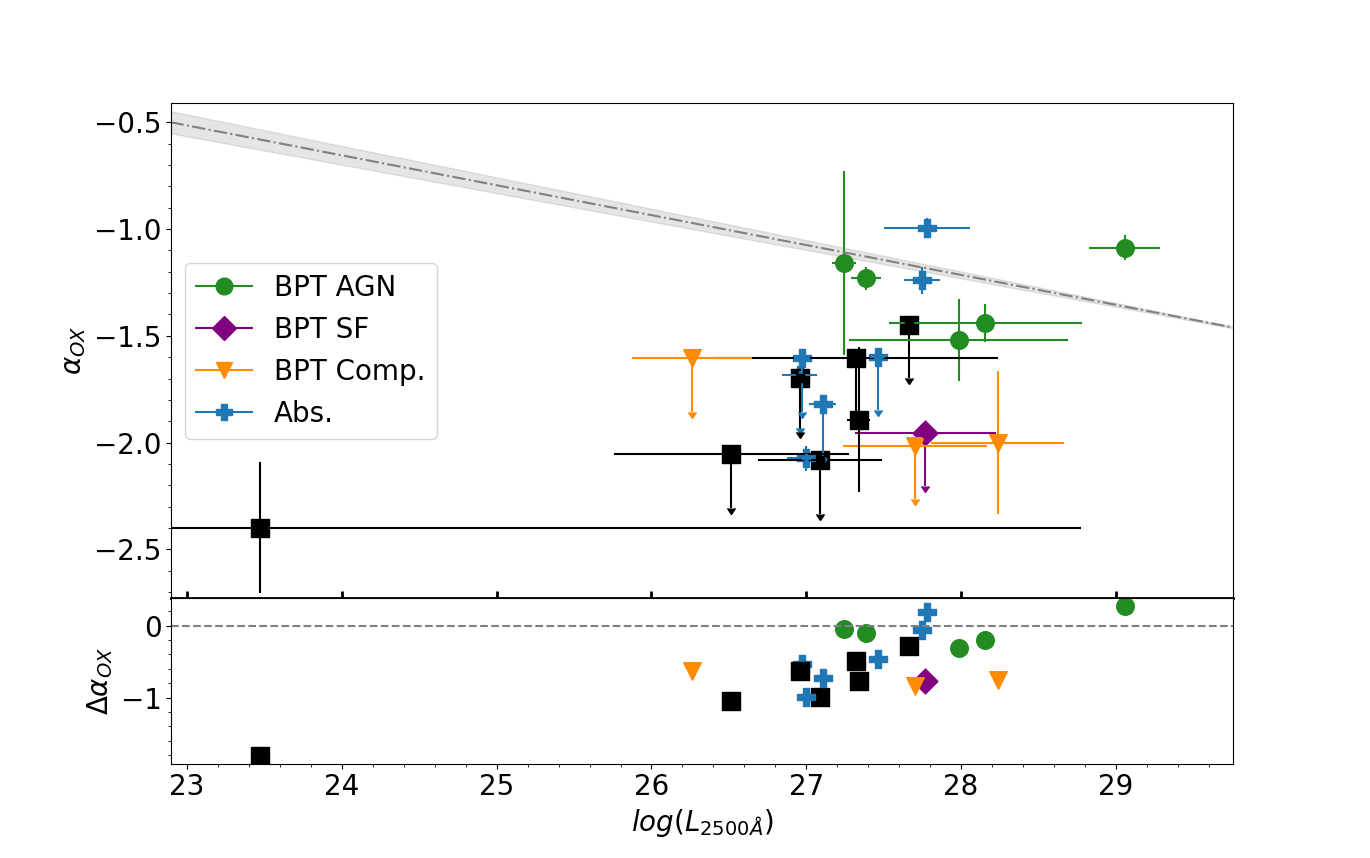}
  \caption{Comparison of $\alpha_{\text{OX}}$ to monochromatic luminosity at 2500\AA. The relation derived for quasars from \cite{just2007} is plotted as the grey line. Residuals from the \cite{just2007} relation are given in the bottom panel. The BPT AGN have the tightest fit to this relation. Marker shapes and colors are the same as those defined for Figure \ref{fig: L_x vs SFR}.}
  \label{fig: Just07 relation}
\end{figure*}

\cite{just2007} investigated the correlation of the X-ray-to-optical slope $\alpha_{\text{OX}}$ to monochromatic luminosity density $l_{2500 \textrm{\AA}}$ for luminous Quasars in the Sloan Digital Sky Survey. They found a clear dependence of $\alpha_{\text{OX}}$ on this UV luminosity and no significant change in the $\alpha_{\text{OX}}$ over the redshift range of $1.5 \leq z \leq 4.5$. We explore whether our sample also follows the $\alpha_{\text{OX}}$ relation in Figure \ref{fig: Just07 relation}. Points are identified by their location on the BPT diagram. Residuals to the \cite{just2007} relation are given in the bottom panel.

Inspection of Figure \ref{fig: Just07 relation} shows that the sub-population with optical spectroscopic AGN signatures best agree with the \cite{just2007} relation, whereas the BPT composite and star forming galaxies, absorption line systems, and those with no optical spectroscopy deviate more significantly from this relation.

As hard X-rays of AGN come from the innermost region of the accretion disk, deviations below the relationship from \cite{just2007} suggest the presence of obscuring material. This obscuring material can not only effect these potential type-II AGNs, but also low-luminosity type-I's. This obscuring gas could be highly ionized, having no significant dust grains due to the heating from the accretion outflow. High resolution spectroscopy of less luminous sources, such as NSA 631492 and 631480, would help decipher between this objects being a type-II or a low luminosity type-I AGN. We investigate the effects of dust further in Section \ref{subsubsec: dust obscuration}.

The deviation from $\alpha_{\text{OX}}$ for these non-spectroscopic AGNs could also suggest a low ionization parameter, $\xi$, of its accretion disk.

We note that some objects with absorption features in their optical spectrum have a closer correspondence than other non-BPT AGN. With optical absorption features suggesting a predominantly older stellar population, the bright X-ray emission of these sources paired with their UV variability support the need for additional multi-wavelength observations for these objects to fully understand the processes at work.

\subsubsection{Dust Obscuration} \label{subsubsec: dust obscuration}

We investigate the potential effect of dust obscuration from the host galaxy on the spectroscopic signatures of our sample. Narrow emission lines are known to to be attenuated by dust (see \cite{hickox2018} and references therein). \cite{rigby2006} find that dust obscuration from the host galaxy is a likely cause of optically-dull emission line ratios. 

\begin{figure*}
    \centering
    \includegraphics[width=0.8\textwidth]{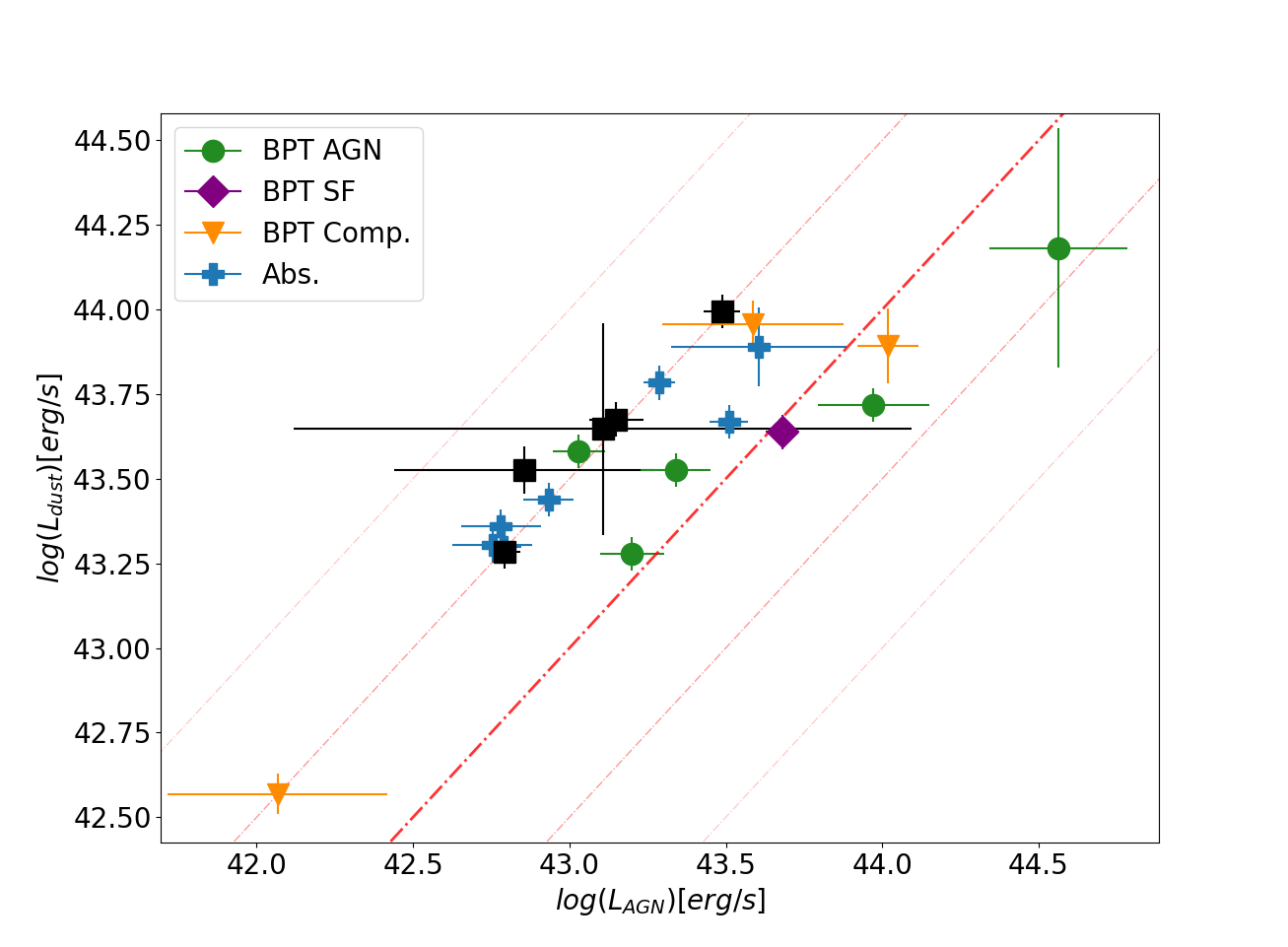}
    \caption{Plot of AGN to dust luminosity for our sample. Line are defined the same as those in Figure \ref{fig:stellar_relation}. Again, points for NSA 631492 and 631480 are left off as they exist in the far left, less luminous regime, compared to the rest of the sample. Marker shapes and colors are the same as those defined for Figure \ref{fig: L_x vs SFR}. }
    \label{fig:dust_relation}
\end{figure*}

For our sample, we find that the dust-to-AGN luminosity plot of Figure \ref{fig:dust_relation} separates these objects similarly to the distinction of Figure \ref{fig:stellar_relation}, as the absorption line dominated and non-BPT AGN have higher dust luminosities than the rest of the sample. However, there is less of a spread in dust lumionsities than is seen for stellar luminosities. 

\subsubsection{Excess Mid-Infrared Luminosity} \label{subsubsec: midIR X-ray relation}

Previous works have used the relationship between mid-IR and X-ray luminosity to explore the accretion physics and dust structure of luminous AGN. \cite{gandhi2009} established a tight linear relation between mid-IR and X-ray luminosity for local Seyfert galaxies. \cite{rigby2006} investigates the mid-IR emission of their sample, finding that their optically dull X-ray AGN have mid-IR emission similar to Seyfert galaxies. Normal AGN optical/UV continua is needed for Seyferts to produce their mid-IR luminosity, thus this rules out a lacking optical/UV continuum as the reason for their optical dullness. 

We use the luminosity at $6\mu$m to investigate the mid-IR regime of their spectrum. This should also help separate the effects of dust from host galaxy and torus components, as mid-IR emission from AGN is produced by torus dust being heated by UV photons.

We investigate the scaling between mid-IR luminosity at $6\mu$m to obscuration-corrected X-ray luminosity, as discussed in \cite{chen2017}. This relation is constructed from a Levenberg-Marquardt minimized $\chi^2$ fitting to AGN populations from the Bo\"otes, XMM-COSMOS, XRT-SDSS, and XXL-N samples assuming a bi-linear, broken power law fit. We note that the "breaking luminosity" of this relation is at $L_{6 \mu m} = 44.79$ erg s$^{-1}$, which is brighter than our most luminous object at $6\mu$m.

Additionally, \cite{fiore2009} and \cite{gandhi2009} have established  an intrinsic mid-IR to X-ray relationships. \cite{gandhi2009} used high resolution photometry of local Seyfert galaxies to find that they follow the same correlation of Compton-thick AGN. The work of \cite{fiore2009} searched for highly obscured, Compton-thick AGN to estimate the missing fraction of AGN  that should be observed in these systems. In doing so, they estimate unobscured X-ray luminosities for source that they originally detect in the mid-IR.

\begin{figure*}
  \centering
    \includegraphics[width=0.8\textwidth]{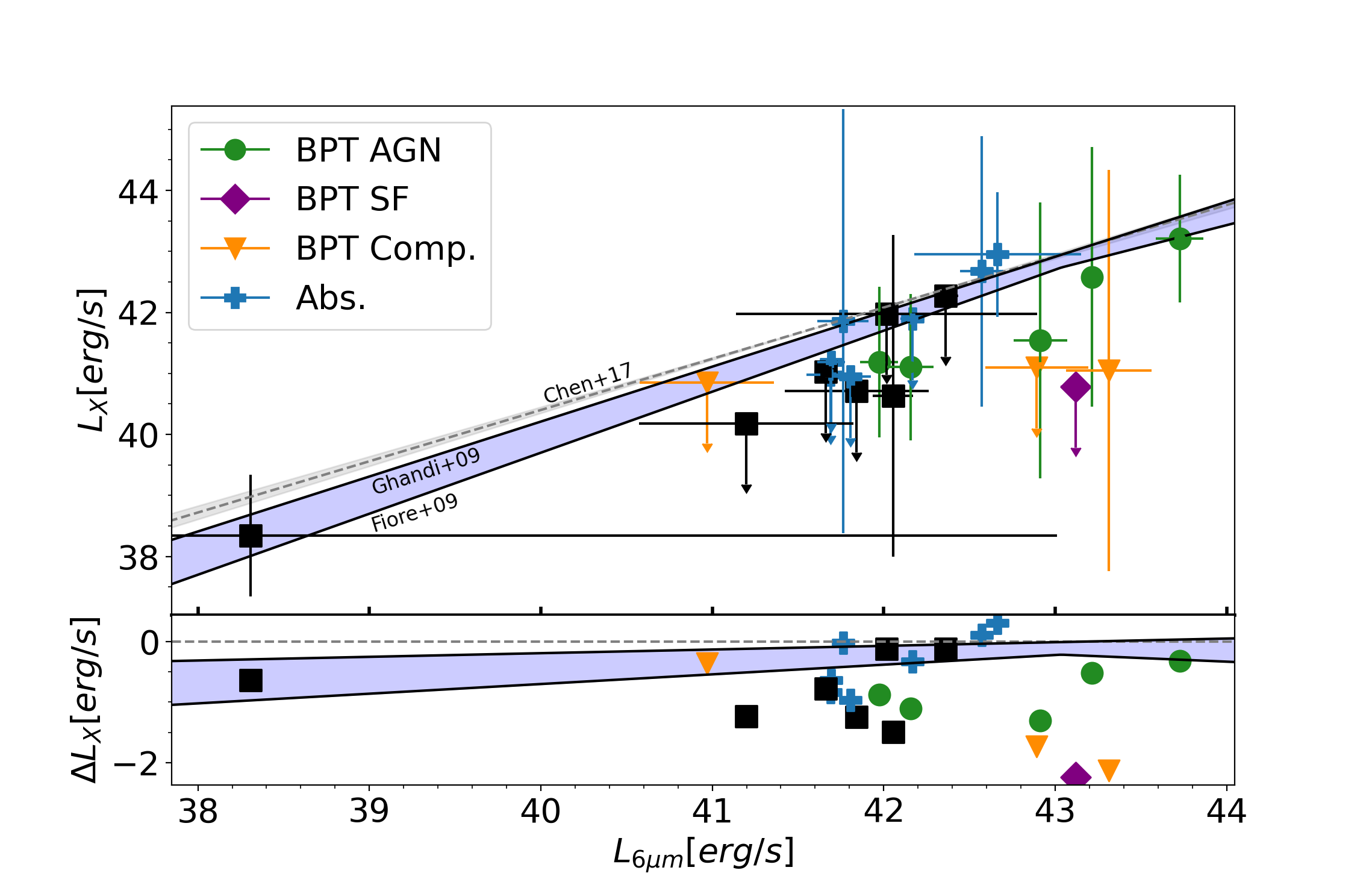}
  \caption{Plot of  2-10 keV luminosity to luminosity at $6 \mu$m. Grey line and shaded region show the empirical relation and its uncertainty from \cite{chen2017}. The intrinsic relationships of \cite{fiore2009} and \cite{gandhi2009} are given as solid black lines, with the blue area shaded between representing a range of possible intrinsic and unabsorbed X-ray luminosity for given mid-IR power. Residuals relative to the \cite{chen2017} relation are again given on the bottom panel. Marker shapes and colors are the same as those defined for Figure \ref{fig: L_x vs SFR}. }
  \label{fig: L_midIR-L_xray relation}
\end{figure*}

We plot this relation against our variable X-ray AGN candidates in Figure \ref{fig: L_midIR-L_xray relation}. The luminosity at $6\mu$m is estimated from our SED models. We find some of the BPT AGN and absorption line dominated systems to be generally consistent with the X-ray to mid-IR relation. This suggests that at least some of the optically dull AGN in our sample have typical AGN optical/UV continuum and torus dust emission.

On the other hand, some of the objects in our sample do have excess mid-IR emission compared to their X-ray luminosity. These include four spectroscopic AGN, two of the absorption line systems, composite objects and some objects without optical spectroscopy. These objects could  have excess dust near the nucleus, potentially explaining the optical dullness in some cases. As mid-IR emission can be an indirect tracer of AGN power \citep{lansbury2015}, this can point to the effect of obscuration within these systems. X-ray absorption causes a steeper slope in the fitting, thus causing some of our spectroscopic AGN to fall below these literary relations.

\section{Conclusions} \label{sec:conclusions}

In this work we measured X-ray emission for 23 UV variability-selected AGN candidates from \cite{wasleske2022} using archival CXO and XMM observations. We detect X-ray emission from 11 of these objects. Then, we constructed spectral energy distribution models using X-CIGALE \citep{yang2020} to measure contributions from dust, stars, and gas across the electromagnetic spectrum. We then explored reasons as to why this sample displays a diversity of optical spectroscopic signatures. We summarize our main findings below.

\begin{itemize}
    \item We measure X-ray emission for 23 sources originally selected as AGN candidates from their UV variability \citep{wasleske2022}. 11 of the 23 are detected.

    \item We compare our observed X-ray luminosities to those expected due to star formation. We find that 10/11 detected galaxies have X-ray luminosities more than 0.6 dex above the \cite{ranalli2003} $L_{X-ray}$-SFR relationship (this is the criteria used for AGN selection in \citealt{agostino2022}). This suggests that these systems are bona fide AGN. The upper limits for the remaining objects are not constraining enough to rule out the presence of AGN.

    \item Three sources, NSA 64146, 64272, and 64286, have detections in both CXO and XMM observations. We argue that the consistency of the X-ray emission over years long timescales is more consistent with AGN rather than a TDE as the possible cause for the X-ray emission and their significant UV variability.

    \item One source, NSA 208625, has bright X-ray emission in CXO observations taken in 2005 and a non-detection in XMM observations taken in 2015. The source faded by at least an order of magnitude over the decade. This could suggest a changing state AGN or a TDE in this galaxy. 

    \item Our confirmed AGN show a diversity of optical spectroscopic properties. We use results from SED modeling to study whether stellar or dust emission could impact the optical spectroscopic properties. We find:
        \begin{itemize}
            \item The absorption line dominated systems generally have a higher ratio of stellar-to-AGN luminosities. This is consistent with a scenario where host galaxy light dominates the optical spectrum.

            \item The BPT AGN are in general good agreement with the relationship of $\alpha_{\text{OX}}$ to luminosity at 2500\AA\ from \citep{just2007}. Other objects show large deviations from this relationship, suggesting possible physical differences from luminous quasars. Obscuring material blocking X-ray emission from the innermost part of the AGN's accretion disk could be responsible for the non-spectroscopic AGN deviating from the relation.

            \item We find excess dust luminosity in those objects in our sample lacking optical spectroscopic signatures. 
            
            \item We also find excess mid-IR emission for some objects in our sample, possibly from the heating of dust or host galaxy contamination.
        \end{itemize}

\end{itemize}

 Ultimately, there is no one clear scenario that explains the diversity of optical spectroscopic properties in this sample. Dilution of AGN signatures by star formation, galaxy-wide dust extinction, and nuclear dust extinction are likely all at play to different degrees in different galaxies. Spatially resolved spectroscopy and more detailed spectral fitting would help to further understand the diversity of properties in this interesting sample.
         
The variability selection of candidates done in \cite{wasleske2022} was complemented by spectroscopic analysis, IR AGN selection and low-resolution SED modeling without X-ray values. This work has expanded on that analysis by including X-ray information and more detailed SED modeling. The results here clearly demonstrate that variability is a robust AGN selection technique, identifying objects that would not have been selected based on their optical spectroscopic properties.

\section{Acknowledgments} \label{sec:acknowledgments}

We thank Erin Kimbro for helpful comments.

This research has made use of data obtained from the 4XMM XMM-Newton serendipitous source catalog compiled by the 10 institutes of the XMM-Newton Survey Science Centre selected by ESA.


\bibliography{ewasleske}{}
\bibliographystyle{aasjournal}

\appendix

\restartappendixnumbering
\section{Example SED Models}
Below we give examples of SED modeling of \texttt{X-CIGALE} for NSA 28616 and NSA 64129. Models for Stellar attenuated, Stellar unattenuated, Nebular, Dust, and AGN emission are given by color yellow, blue, green, red, and orange respectively. Total model is shown in black with model fluxes as red dots and observed fluxes as purple circles.

Both object's SED modeling has significant AGN contributions in the IR. The model fro NSA 28616 shows extinction of the AGN emission at higher energies, but the sole contribution to the X-ray portion as the total black model line is sitting on top of the orange AGN emission line.  The model for NSA 64129 shows heavy extinction in the ultraviolet.

\begin{figure*}[hbt!]
    \centering
    \subfigure{\includegraphics[scale=0.16225]{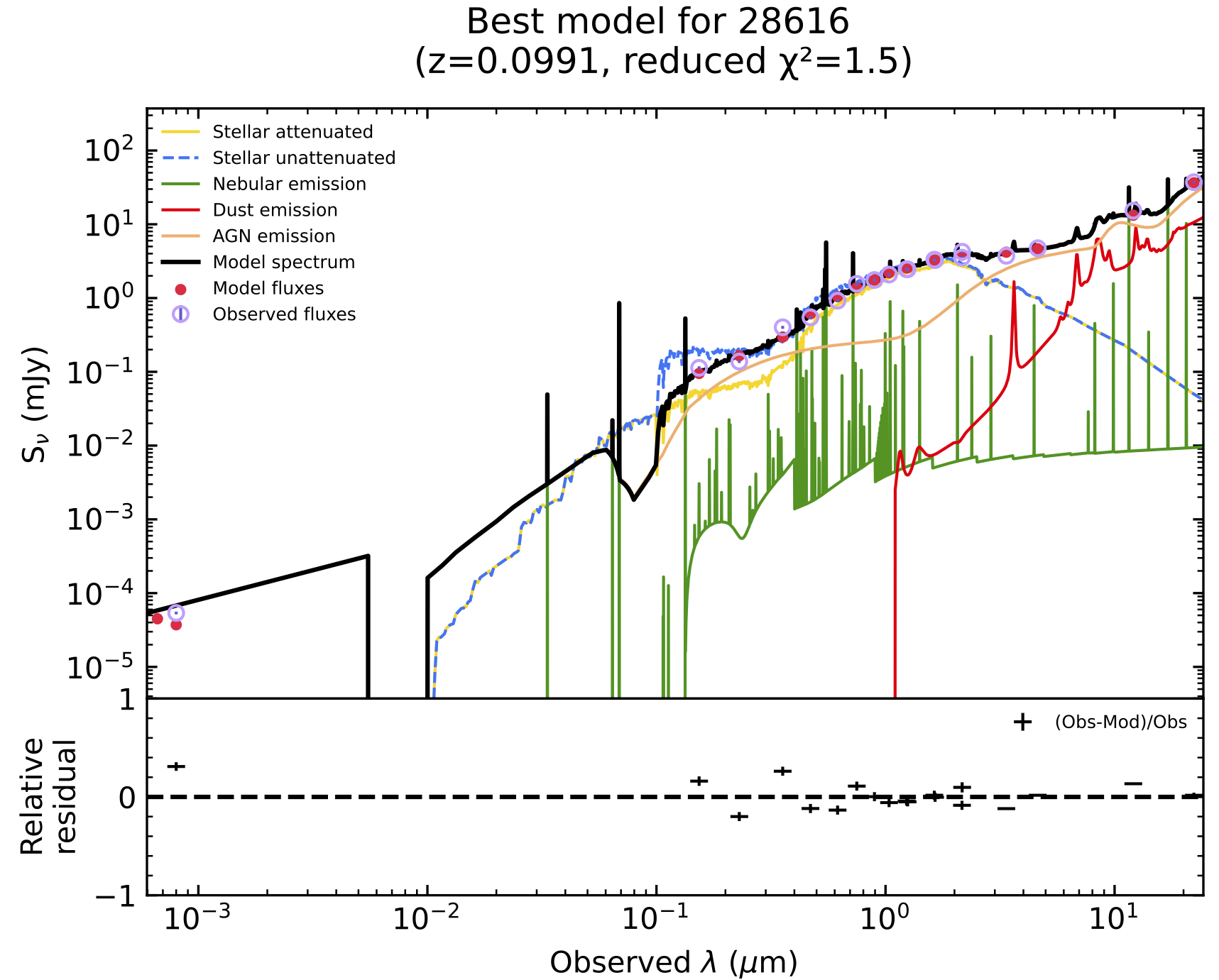}}
    \hspace{0.25em}
    \subfigure{\includegraphics[scale=0.16225]{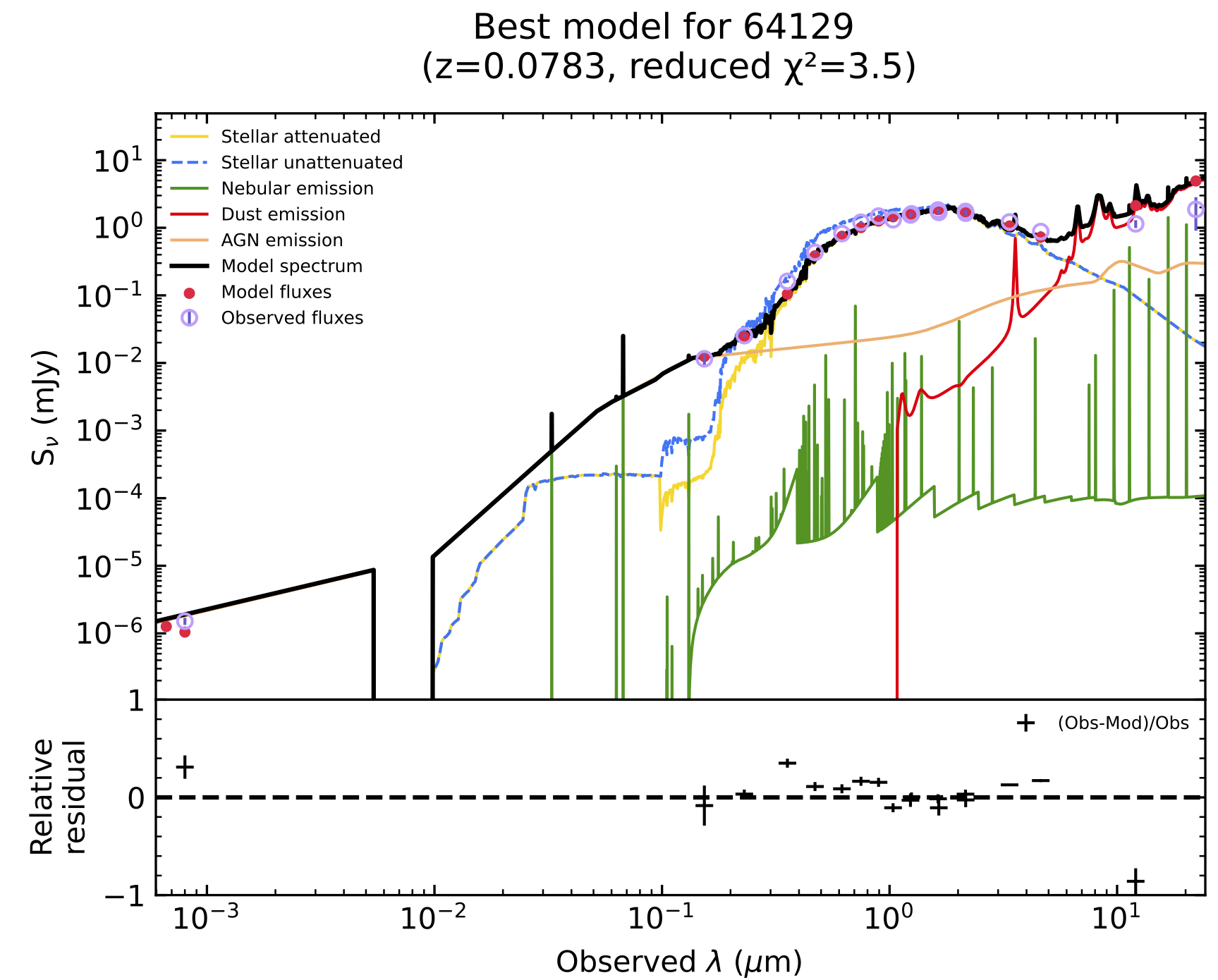}}
    \caption{SED Models of NSA 28616 (\emph{Top}) and NSA 64129 (\emph{Bottom}). } 
    \label{fig:SED_model_exs}
\end{figure*}

\end{document}

%% file: table1.tex
\begin{table*}[ht]
    \centering
    \textbf{X-ray Sample\\}
    \begin{tabular}{c c c c c c c c c}
\hline
NSA ID  & z & $\log ( \text{M}_{*} / \text{M}_\odot )$  & BPT Class & \multicolumn{3}{|c|}{CXO} & \multicolumn{2}{|c|}{XMM}  \\
\hline
    &   &   &   &  Counts &  $n_H$   &$L_{0.5-7.0 \rm{keV}}$    & $n_H$    & $L_{0.5-7.0 \rm{keV}}$  \\
    &   &   &   &        &  [$10^{22} \text{cm}^{-2}$]    & [erg/s]    &  [$10^{22} \text{cm}^{-2}$]   &  [erg/s] \\
\hline
28616 & 0.0991 & 10.89 & AGN           & -- & -- & -- &    $0.106$ & $ 2.8^{+0.1}_{-0.09} \times 10^{43}$ \\
28810 & 0.1394 & 11.24 &    None       & $\leq 15 \pm 4$& -- &$< 8.6 \times 10^{41}$ &  -- &$< 7.7 \times 10^{41}$ \\ 
64129 & 0.0783 & 10.53 & AGN           &  -- & -- & -- &   $0.096$   & $ 5.0^{+0.5}_{-0.5} \times 10^{41}$ \\
64145 & 0.1030 & 10.59 &    Absorption & $1397 \pm 37$ & $2.230$ &  $(3.6 \pm 0.7) \times 10^{42}$ & $0.045$  & $(1.2 \pm 0.3) \times 10^ {42}$ \\
64258 & 0.1084 & 10.74 &	Starburst  &  $\leq 8 \pm 2$ & -- & $< 7.3 \times 10^{39}$  & --  & $< 2.1 \times 10^{41}$ \\ 
64266 & 0.0805 & 10.68 &    None       &  $\leq 12 \pm 4$ & --  & $< 6.5 \times 10^{39}$ &  --  &$< 1.5 \times 10^{41}$ \\ 
64272 &	0.1329 & 10.89 &	AGN	       &  $554 \pm 24$& -- & $(4.6 \pm 0.3) \times 10^{42}$ & $0.107$ & $(9.7 \pm 0.2) \times 10^{42}$ \\
64286 &	0.0234 & 10.67 &	AGN	       &  $97 \pm 10$ & $1.667$ & $2.7 ^{+3.4}_{-0.7}\times 10^{41} $ & $0.036$   & $(5.3 \pm 0.2) \times 10^{41}$ \\
205160 & 0.1396 & 10.52  &  None       & -- & -- & --&   --&$< 1.3 \times 10^{42}$ \\ 
208465 & 0.0765 & 11.61 & Absorption   & -- & -- & -- &          $0.055$  & $(2.4 \pm 0.03) \times 10^{43}$ \\
208625 & 0.0732 & 10.85 & Absorption   & $503 \pm 22$ & --  & $1.4^{+0.6}_{-0.7} \times 10^{42}$ &        --  & $< 1.4 \times 10^{41}$ \\ 
208662 & 0.1122 & 10.98 & Composite    & $455 \pm 21$ & $2.49$  &$8.4^{+10.1}_{-7.0} \times 10^{40}$ &                   --  &  $< 1.2 \times 10^{42}$ \\  
208702 & 0.0448 & 9.61  &  Composite   &  -- & -- & -- &               --  &   $< 9.9 \times 10^{40}$ \\ 
259478 & 0.0784 & 10.84 & Composite    &  -- & -- & --  &                     --  & $< 1.7 \times 10^{41}$ \\ 
259789 & 0.1063 & 10.76 & None         & $10038 \pm 100$ &     $0.292$  &   $4.9 ^{+3.1}_{-2.5} \times 10^{40}$ &             --  &   --  \\
259880 & 0.0762 & 10.65	& Absorption   &  -- & -- &  --  &              --  &   $< 1.3 \times 10^{41}$ \\ 
259895 & 0.0834 & 10.11 & AGN          & -- & -- &   --  &              $0.226$    & $(2.4 \pm 0.2) \times 10^{42}$ \\
259919 & 0.0781	& 10.52	& Absorption   & -- & -- &  --  &                 --  &  $< 1.2 \times 10^{41}$ \\ 
260221 & 0.0742	& 10.73 & Absorption   & -- & -- &  --  &               --  &    $< 2.1 \times 10^{41}$ \\ 
260241 & 0.0743	& 10.51	& None	       & -- & -- &  --  &               --  &    $< 1.5 \times 10^{41}$ \\ 
613397 & 0.1274 & 11.54 & Absorption   & $\leq 9 \pm 3$ &  --  & $< 3.3 \times 10^{41}$ &          --  &   --\\
631480 & 0.0442 & 7.74 & None          & $\leq 5 \pm 3$ &   --  &$< 9.8 \times 10^{38}$  &          --  &   $< 3.6 \times 10^{40}$ \\ 
631492 & 0.0119 & 6.89 & None	       & $25 \pm 5$ &   --  &   $7.8^{+10.8}_{-6.7} \times 10^{37}$  &      --  &   $< 8.0 \times 10^{39}$ \\

\hline

    \end{tabular}
    \caption{ Our sample from \cite{wasleske2022} of UV variable AGN candidates found in X-ray surveys. Counts for CXO detections are calculated by their $0.5-7.0 \rm{keV}$ count rate multiplied by the exposure time, with Poissonian errors given. Their BPT classifications are given as ``BPT Class", with some spectrum dominated by absorption lines. Those galaxies with a BPT Class of 'None' do not have available spectra that can be properly fitted and analyzed. Intrinsic absorption $n_H$ values given for detections in CXO and XMM, based on the methods described in Section \ref{subsec: ciao reduction} and \ref{subsec: source catalog}. }
    \label{tab:lum table}
\end{table*}